\documentclass[manuscript]{acmart}
\usepackage{wrapfig}
\usepackage{xcolor}
\usepackage{caption}
\usepackage{subcaption}
\usepackage{lipsum} 

\AtBeginDocument{%
  \providecommand\BibTeX{{%
    \normalfont B\kern-0.5em{\scshape i\kern-0.25em b}\kern-0.8em\TeX}}}

\copyrightyear{2022}
\acmYear{2022}
\setcopyright{rightsretained}
\acmConference[RecSys '22]{Sixteenth ACM Conference on Recommender
Systems}{September 18--23, 2022}{Seattle, WA, USA}
\acmBooktitle{Sixteenth ACM Conference on Recommender Systems
(RecSys '22), September 18--23, 2022, Seattle, WA, USA}
\acmDOI{10.1145/3523227.3547390}
\acmISBN{978-1-4503-9278-5/22/09}

\begin{document}

\title{An Incremental Learning framework for Large-scale CTR Prediction}


\author{Petros Katsileros}
\orcid{0000-0002-8682-2560}

\author{Nikiforos Mandilaras}
\orcid{0000-0002-3014-9813}

\author{Dimitrios Mallis}
\orcid{0000-0003-0962-3477}

\author{Vassilis Pitsikalis}
\orcid{0000-0002-1593-7491}

\author{Stavros Theodorakis}
\orcid{0000-0002-8282-3558}

\affiliation{%
  \institution{Deeplab}
  \city{Athens}
  \country{Greece}
}

\affiliation{%
  \institution{Taboola.com}
  \city{Tel Aviv}
  \country{Israel}
}

\author{Gil Chamiel}
\orcid{0000-0001-7040-173X}

\affiliation{%
  \institution{Taboola.com}
  \city{Tel Aviv}
  \country{Israel}
}

\renewcommand{\shortauthors}{P. Katsileros, N. Mandilaras et. al}

\begin{abstract}

In this work we introduce an incremental learning framework for Click-Through-Rate (CTR) prediction and demonstrate its effectiveness for Taboola's massive-scale recommendation service. Our approach enables rapid capture of emerging trends through warm-starting from previously deployed models and fine tuning on ``fresh'' data only. Past knowledge is maintained via a teacher-student paradigm, where the teacher acts as a distillation technique, mitigating the catastrophic forgetting phenomenon. Our incremental learning framework enables significantly faster training and deployment cycles ($x12$ speedup). We demonstrate a consistent Revenue Per Mille (RPM) lift over multiple traffic segments and a significant CTR increase on newly introduced items.

\end{abstract}



\maketitle

\section{Introduction}
Over parameterised neural networks have recently demonstrated strong Click-Through-Rate (CTR) prediction performance for large-scale ad recommendations \cite{widedeep,deepfm}. To effectively model user preferences and intent, such models generally train on historical data collected over the span of several days (or even weeks) and commonly require long training times. On the other hand, model freshness constitutes an important factor for effective recommendations. Given the dynamic nature of the deployment environment, data distributions can vary significantly with passing time, due to various factors like seasonality, the addition of new items, etc. Thus, prediction accuracy clearly degrades with increased delay between training and inference phases \cite{practicalLess}.

As a leading content recommendation service, Taboola serves more than a billion requests to millions of unique users each day. Under this industrial setting, hundreds of CTR prediction models are deployed daily (on distinct segments of incoming traffic). Each model is trained from scratch for several hours on historical user impressions (collected over the previous two weeks). Thus, long training times introduce challenges with regard to \textit{(1)} model freshness as new trends constantly emerge (that are not captured by historical data) and \textit{(2)} the scale of required computational resources.

To address these challenges, we introduce Taboola's incremental learning framework for CTR prediction. Instead of training each model from scratch on historical user impressions, we opt for initialising from past-deployed models through a warm-start step. New models are trained on fresh data only, which comprises only a small fraction of the original dataset. During training, we also employ a teacher-student paradigm, where the teacher (trained daily) acts as an implicit regularizer, enabling the student to maintain previously acquired knowledge. The presented approach drastically reduces the required training time, thus allowing the deployment of “fresher” models with less computational requirements. We demonstrate a significant CTR increase on newly introduced items and overall improvement in recommendation performance, measured in terms of Revenue Per Mille (RPM) (lift of more than 0.5\%). Note that given the massive scale of Taboola's recommendation service, even a minor RPM increase can be significant.

\section{Method}

This section discusses the different components of our incremental learning framework for the efficient training of ``fresh'' CTR prediction models.

\subsection{Warm Start}

Warm-starting refers to the common practice of initialising the weights of a neural network from a pre computed model \cite{ash2020warmstarting}. This is in contrast to providing a fresh initialisation, also referred to as training from scratch. Given a strong initialisation, warm-started models can be then fine tuned in related domains and achieve strong performance, particularly when data for the target domain are scarce. Warm-start has been shown to improve model robustness and uncertainty estimates in \cite{robustUncert}. Training time is also drastically reduced since fine tuning commonly requires fewer training iterations. 

\subsection{Knowledge Distillation}

The teacher-student architecture employed in this work is based on the popular Knowledge Distillation (KD) framework of \cite{hinton2015distilling}. KD is an efficient type of model compression where knowledge from a larger teacher model is distilled into a smaller student. The student model is trained to predict the true target labels and match the soft targets provided by the larger teacher. Soft targets are the output of a softmax layer that converts the teacher’s logits into class probabilities. Commonly, a temperature parameter T is also used (on the softmax calculation) to produce a softer probability distribution over classes. The student network is trained to minimise the following objective:

\begin{equation} \label{eq:1}
    \mathcal{L}_{S} =  \mathcal{L}_{CE}(y,  \hat{y}_S) +  \alpha \cdot \mathcal{L}_{CE}(\hat{y}_T, \hat{y}_S)
\end{equation}
where $\mathcal{L}_{CE}$ is the standard cross entropy loss, $\hat{y}_S$ and $\hat{y}_T$ are the predicted probabilities of the student and teacher models respectively, $y$ is the ground truth click label and $\alpha$ is a scaling parameter. Our approach also relates to recent methods for incremental learning where knowledge distillation is used to prevent catastrophic forgetting for image classification \cite{Li2018LearningWF, icarl}.

\subsection{Warm Start \& Knowledge Distillation for CTR prediction in Taboola}

In this subsection, we describe our incremental learning framework. A teacher is trained daily (from scratch) on distinct partitions of the data. Note that in Taboola, we train distinct models for different segments, each covering multiple business and modelling aspects like language, categories, publishers and other supply and demand parameters.

Each time a new student is trained, model weights are initialised from the teacher network. Note that in contrast to \cite{hinton2015distilling}, KD is not used for model compression but for diminishing catastrophic forgetting ~\cite{empiricInv}. Thus, we opt for student and teacher models of equal size. In practice, the number of parameters between teacher and student differs as new rows are added to the student's embedding tables (due to the introduction of new categorical variables for the new items). We opt for simply learning new embeddings from scratch as their weights cannot be initialised from the teacher model.

\begin{figure*}
    \centering
    \includegraphics[width=\linewidth]{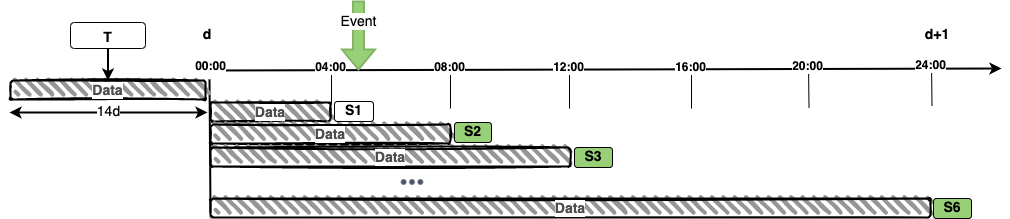}
    \Description[Illustration of the Incremental framework deployments]{The Figure depicts training timelines for teacher and student models throughout the day. It also illustrates the time span, training data of each student model are capturing. A single teacher is used to initialise and regularise multiple students through the day (one every 4 hours). A new event (introduced at 4:15) is rapidly captured by the following student model on 8:00 (within a 4h window).}
    \caption{Illustration of the proposed framework. A single teacher model (T) is used to initialise and regularise multiple students (S1-S6) through Warm-Start and KD. Our framework allows the rapid deployment of fresh models. Notice that a new event (marked with a green arrow) is rapidly captured from the subsequent student model, only a few hours later.}
    \label{fig:novel_trend_fig.png}
\end{figure*}

The student is then fine tuned on fresh data only, collected over a time window spanning a few hours prior to model training. This approach allows the timely modelling of emerging trends on incoming traffic. In the illustration of Fig. \ref{fig:novel_trend_fig.png} for example, a new event or trend will be rapidly captured, with only a few hours of delay (compared to daily model deployments). Moreover, our teacher-student architecture acts as an implicit regulariser that diminishes the effect of catastrophic forgetting (similar to \cite{Li2018LearningWF,icarl} for image classification). The student preserves previously acquired knowledge by predicting the relative probabilities (soft targets) provided by the teacher. This is in contrast to the original setting examined in \cite{hinton2015distilling} where teacher and student are trained under the same data distribution.

\section{Implementation Details}

Each teacher is trained once per day for approximately 2.5 hours on a dataset of $ \approx100M$ samples (user impressions) collected over the last 14 days (see Fig. \ref{fig:novel_trend_fig.png}). Our incremental learning framework enables a significant speedup in fresh model deployment. Student training requires approximately 12 minutes. This is a $12.5x$ speedup compared to model training from scratch with historical data. Note that, to enable Taboola's personalised recommendation service, more than 100 models are daily trained to capture distinct segments of the data or enable internal R\&D experimentation (A/B tests). Thus, such a speedup translates to significant savings in computational resources.

In our production pipeline, we perform student training at 4-hour intervals, leading to 6 deployment cycles per day (see Fig. \ref{fig:novel_trend_fig.png}). Students are fine tuned on $\approx12M$ samples collected from fresh incoming traffic. For implementation efficiency, teacher soft-targets are pre computed as a separate post-operation of the data prepossessing step. In practice, as more traffic becomes available, the number of training samples for students trained later in the day increases. This leads to a negligible training time increase (only a few minutes approximately).

\section{Results}

\begin{wrapfigure}{r}{0.65\linewidth}
     \centering
     \begin{subfigure}[t]{0.32\textwidth}
        \centering
        \includegraphics[width=\linewidth]{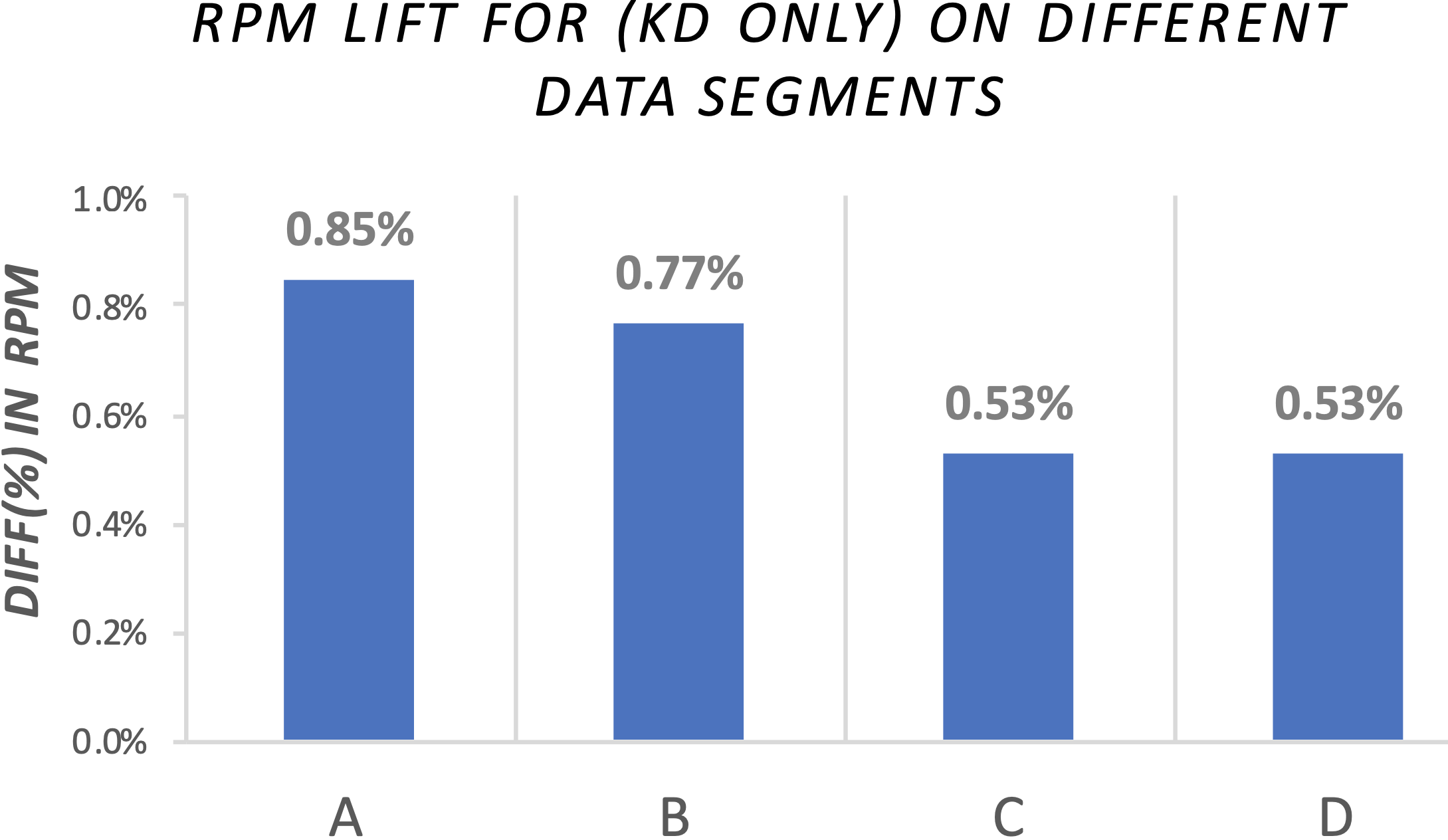}
        \Description[RPM boost using KD in four major data segments.]{Visualization of the RPM impact of KD, applied in four different data segments of our system. Specific RPM values for each data segment are: [Data segment A: +0.85\%, Data segment B: +0.77\%, Data segment C: +0.53\%, Data segment D: +0.53\%].}
        \caption{}
        \label{fig:res1}
     \end{subfigure}
     \hfill
     \begin{subfigure}[t]{0.32\textwidth}
        \centering
        \includegraphics[width=\linewidth]{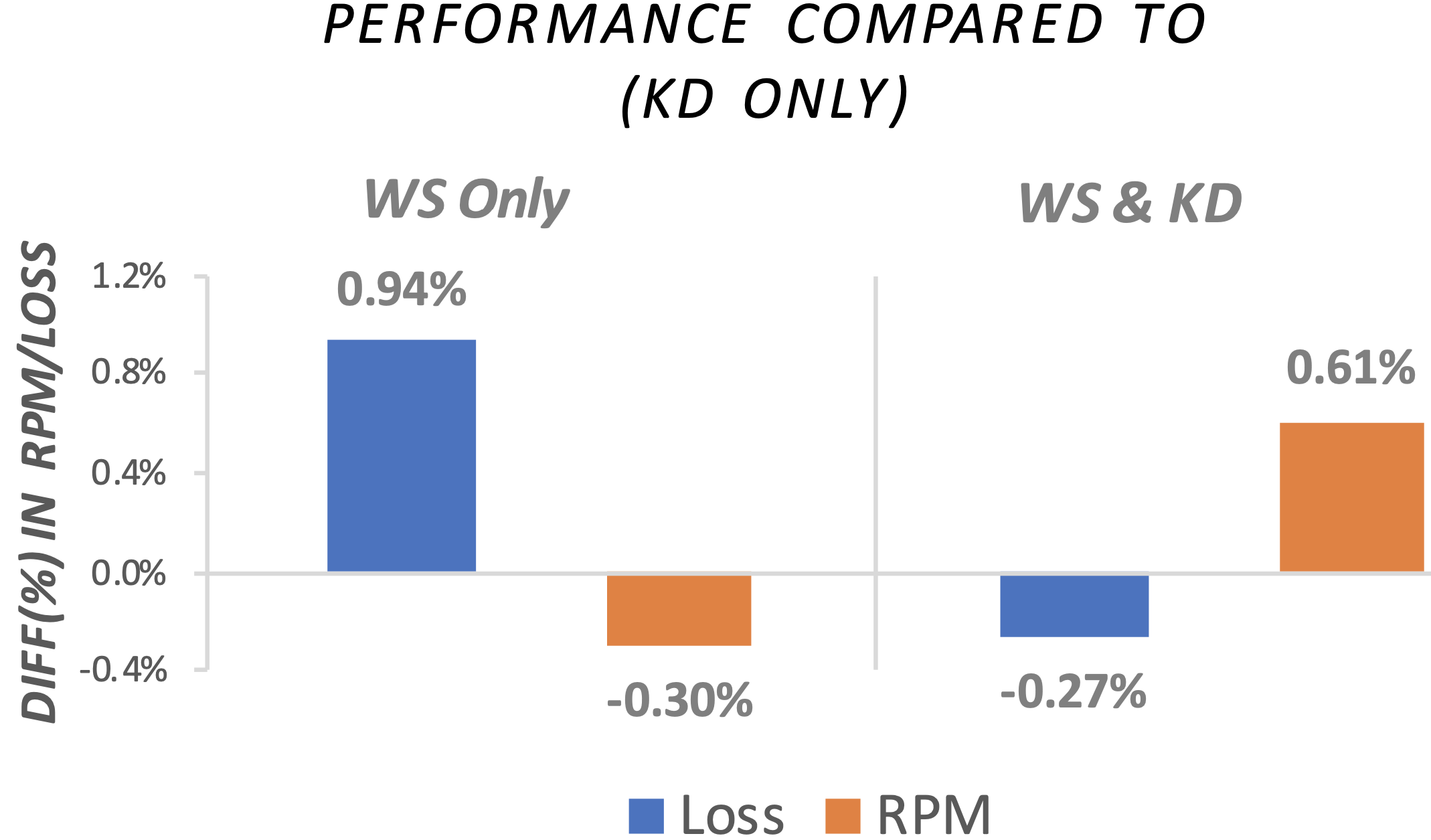}
        \Description[Ablation study for (WS only) and (WS \& KD).]{Visualization of the RPM and Log-loss values between (WS only) and (WS \& KD), applied on a specific data segment and compared with our baseline. For (WS only) we have the following results (percentage differences): [Log-loss: +0.94\%, RPM: -0.3\%] while for the (WS \& KD): [Log-loss: -0.27\%, RPM: +0.61\%].}
        \caption{}
        \label{fig:res2}
     \end{subfigure}
     \hfill
     \vspace{12pt}
     \begin{subfigure}[b]{0.32\textwidth}
        \centering
        \includegraphics[width=\linewidth]{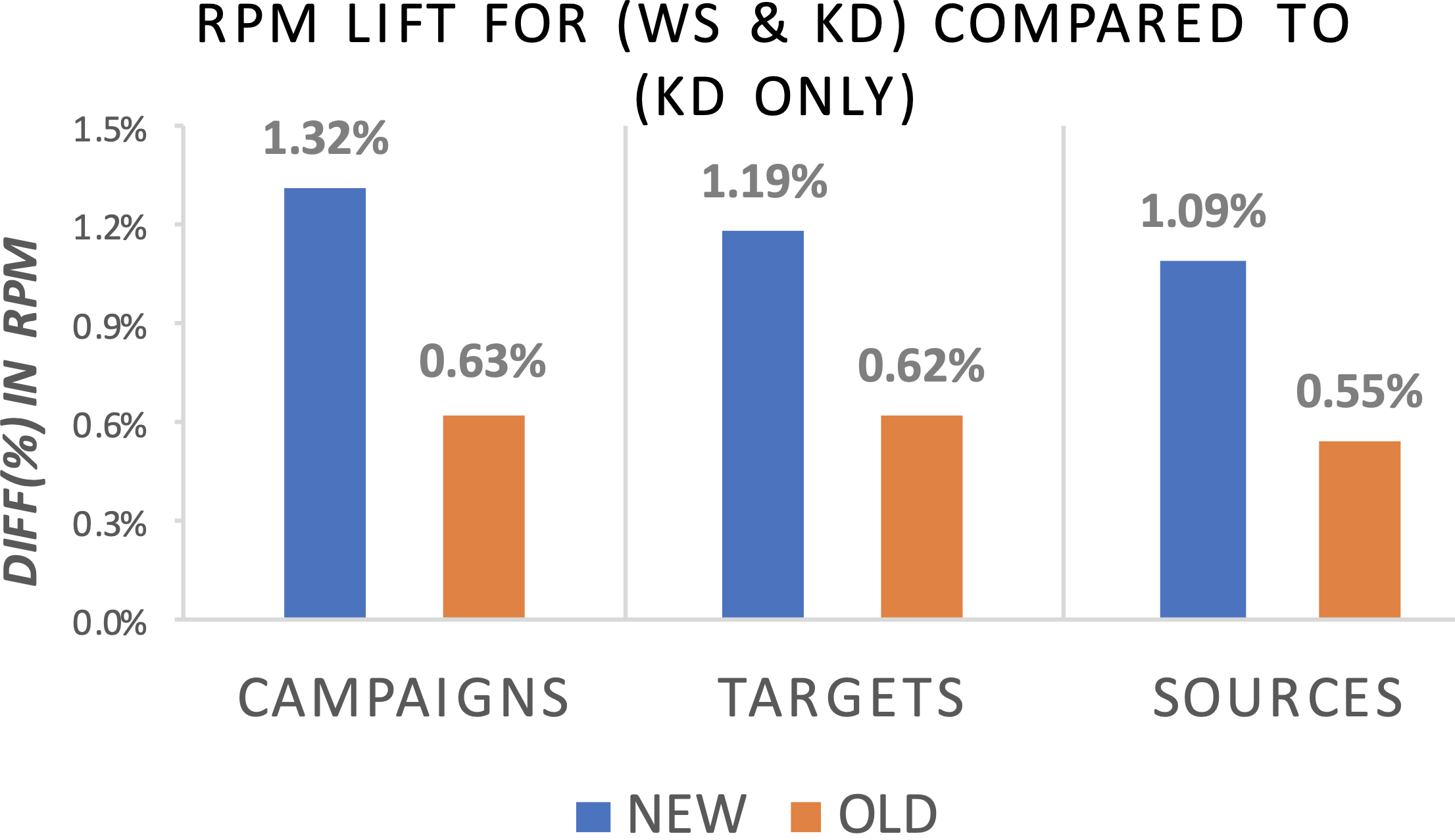}
        \Description[Illustration for the RPM gains of the proposed Incremental learning framework  on ``new`` and ``old`` Campaigns, Targets and Sources.]{Visualization of the RPM impact of the proposed Incremental learning framework on ``new`` and ``old`` Campaigns, Targets and Sources, measured on a particular data segment. We show the following results: [New Campaigns: +1.32\%, Old Campaigns: +0.63\%], [New Targets: +1.19\%, Old Targets: +0.62\%], [New Sources: +1.09\%, Old Sources: +0.55\%].}
        \caption{}
        \label{fig:res3}
     \end{subfigure}
     \hfill
     \begin{subfigure}[b]{0.32\textwidth}
        \centering
        \includegraphics[width=\linewidth]{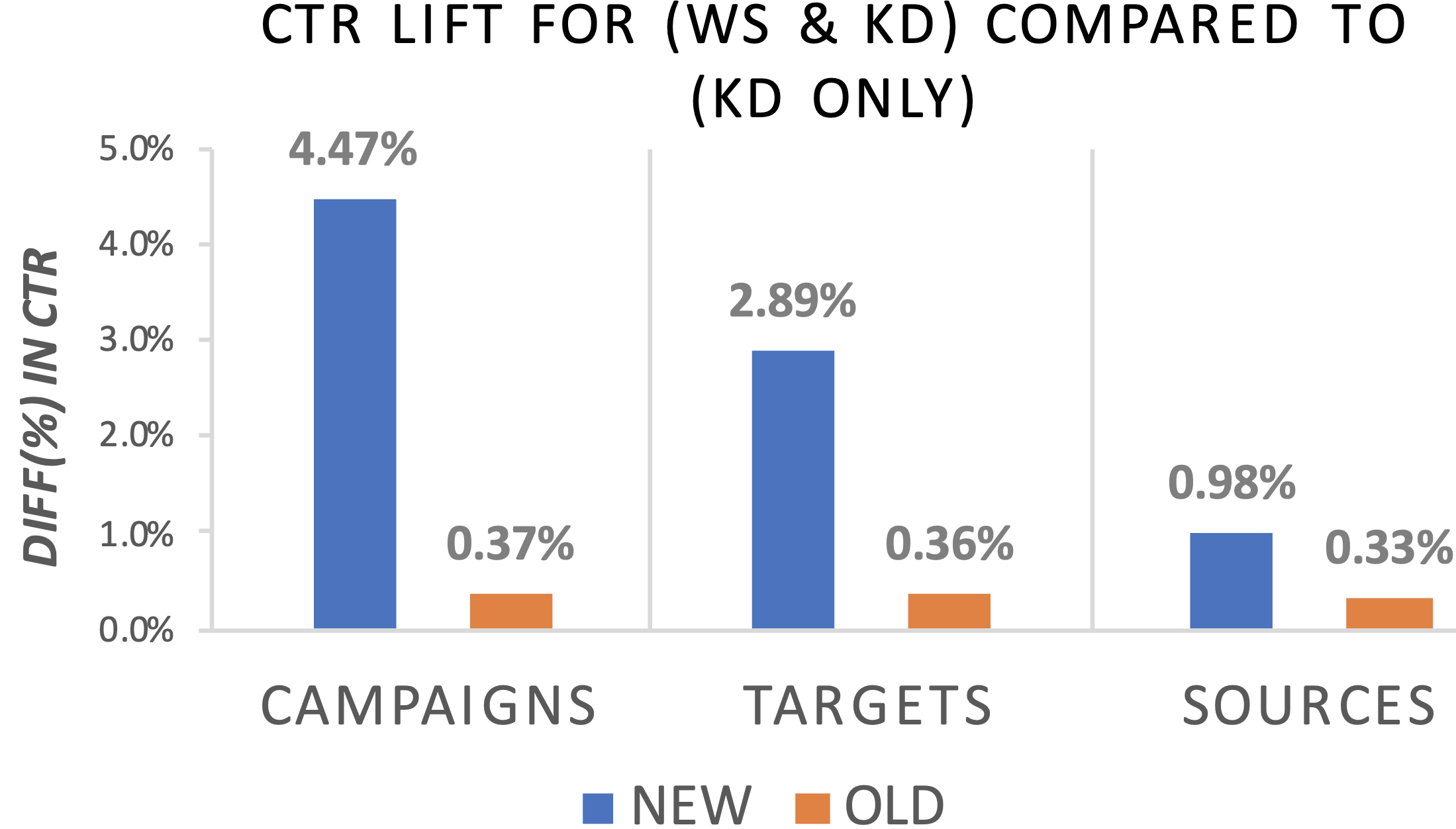}
        \Description[Illustration for the CTR increase of the proposed Incremental learning framework  on ``new`` and ``old`` Campaigns, Targets and Sources.]{Visualization of the CTR impact of the proposed Incremental learning framework on ``new`` and ``old`` Campaigns, Targets and Sources, measured on a particular data segment. We show the following results: [New Campaigns: +4.47\%, Old Campaigns: +0.37\%], [New Targets: +2.89\%, Old Targets: +0.36\%], [New Sources: +0.98\%, Old Sources: +0.33\%].}
        \caption{}
        \label{fig:res4}
     \end{subfigure}
        \caption{Evaluation for the presented framework. \textbf{(a)} RPM lift for (KD only) on four major data segments, compared to (without WS or KD) baseline, \textbf{(b)} Performance comparison for our full incremental learning pipeline (WS \& KD) and simple fine tuned variation (WS only) \textbf{(c,d)} RPM and CTR lift of (WS \& KD) on ``new`` and ``old`` content.}
        \label{fig:three graphs}
\end{wrapfigure}

In this section, we report on the performance of our incremental learning framework for CTR prediction on Taboola's ad recommendation system (measured through A/B tests).
We first evaluate the examined student-teacher training pipeline with an A/B test (lasting $\approx 6$ months) on four major data segments. A student trained from scratch over historical data (KD only) is compared to a baseline model (without WS or KD). Lift in RPM is shown in Fig. \ref{fig:res1}. We observe that our teacher-student pipeline results in a consistent performance increase (ranging from 0.53\% to 0.85\%) in all cases. Following these A/B test results, the KD setting was rolled out as Taboola's default production setup and will be our baseline for the comparisons shown in the remainder of this section.

In Fig. \ref{fig:res2}, we evaluate a model trained with our complete incremental learning framework (WS \& KD). Results for a model trained with warm-start only (WS only) are also shown. We observe that simply fine tuning on the fresh data (WS only), results in performance decrease (RPM regret and increase in Log-loss) compared to the baseline (KD only). In contrast, our incremental learning framework (WS \& KD), leads to consistent performance gains (RPM lift and decreased Log-loss). Warm-start enables rapid training and deployment of new models and our teacher-student pipeline mitigates the effect of catastrophic forgetting as the student effectively maintains previously acquired knowledge. Note that given the scale of Taboola’s content recommendation service, even a small RPM increase (we report a 0.61\% increase on the performed A/B test, lasting $\approx 2$ months) has a significant business impact.

Finally, one of the main benefits of our approach is the timely capture of emerging trends in user preferences and rapid bootstrap of newly available items. To quantify this effect in Fig. \ref{fig:res3}, \ref{fig:res4}, we evaluate the performance  on newly introduced targets, campaigns (usually include multiple targets) and sources (pages where targets are served). Our incremental learning framework (WS \& KD) leads to consistent RPM and CTR increase for newly introduced items (compared to the KD only baseline), demonstrating our methods' ability to capture emerging trends on fresh data.

\section{Conclusion}
We present Taboola's incremental learning framework for CTR prediction. A combination of warm start and teacher-student techniques is utilised to minimise training and deployment cycles and improve model performance. Our solution increases both RPM and CTR of ``newly'' introduced items and boosts the overall system's performance in terms of RPM.

\clearpage

\bibliographystyle{ACM-Reference-Format}
\bibliography{bibliography}

\appendix

\end{document}